\documentclass[12pt]{article}
\usepackage[margin=1in]{geometry}
\usepackage{changepage}
\usepackage[braket, qm]{qcircuit}
\usepackage{graphicx}
\usepackage{subcaption}
\usepackage{booktabs}
\usepackage{amsmath, amssymb, braket, bm}
\usepackage{float}
\usepackage[printonlyused,nolist]{acronym}

\usepackage{listings}
\lstnewenvironment{usagesnippet}{\lstset{xleftmargin=1em,xrightmargin=1em}}{}
\usepackage{upquote}
\usepackage[T1]{fontenc}

\usepackage{algorithm}
\usepackage{algorithmic}

\usepackage{tikz}
\usepackage[hidelinks]{hyperref}
\usepackage{pdflscape}
\hypersetup{
    pdfauthor={Microsoft Quantum},
    pdftitle={Clifford-efficient sparse state preparation for molecular wavefunctions},
    pdfsubject={Sparse quantum state preparation for molecular wavefunctions},
    pdfkeywords={quantum state preparation, sparse quantum states, quantum chemistry, Clifford circuits, finite-field linear algebra}
}
\usepackage{authblk}

\renewcommand{\vec}[1]{\bm{#1}}
\newcommand{\mat}[1]{\mathsf{#1}}
\newcommand{\CNOT}{\operatorname{CX}}
\newcommand{\X}{\operatorname{X}}

\newcommand{\Toffoli}{\operatorname{Toffoli}}

\title{Clifford-efficient sparse state preparation for molecular wavefunctions}

\author{Yingrong~Chen\thanks{Primary author. Other authors listed in alphabetical order. Work by HL, YS, and YY was done while at Microsoft.}}
\author{Nathan~A.~Baker}
\author{Rushi~Gong}
\author{Conrad~S.N.~Johnston}
\author{Brad~Lackey}
\author{Hongbin~Liu}
\author{Sasha~Schmidt}
\author{Yuan~Su}
\author{David~B.~Williams-Young}
\author{Yinuo~Yang}
\affil{Microsoft Quantum, Redmond, WA, USA}

\date{\today}

\begin{document}

\maketitle

\begin{abstract}
Sparse quantum state preparation concerns an $n$-qubit target state that is a superposition of only $d \ll 2^n$ computational basis states.
Existing approaches exploit this sparsity by compressing these $d$ basis states and their amplitudes onto a smaller set of qubits, called the dense register, before expanding the prepared state to the full register.
Rather than relying on the permutation-based compression used in prior work, we exploit affine relationships among the binary configurations over the finite field $\operatorname{GF}(2)$ to reduce both the non-Clifford gate count and the ancillary qubit count.
Invertible affine transformations over $\operatorname{GF}(2)$, comprising Gaussian elimination and all-ones-row removal, first reduce the dense register from $n$ to the rank $r$ using only Clifford gates and no ancillary qubits.
An optional binary encoding stage then trades additional $\Toffoli$ gates and ancillary qubits for further compression to the minimum $\lceil\log_2 d\rceil$ dense qubits needed to represent $d$ distinct configurations.
For chemically relevant wavefunctions, such as those obtained from selected configuration interaction calculations, shared electronic excitation patterns produce many of these affine relationships, enabling substantial Clifford-only compression before binary encoding.
Across the molecular benchmarks, our method requires the fewest ancillary qubits among the evaluated sparse state preparation methods while maintaining comparable non-Clifford gate counts when using binary encoding.
\end{abstract}

\tableofcontents

\begin{acronym}
    \setlength{\itemsep}{0pt}
    \setlength{\parskip}{0pt}
    \setlength{\parsep}{0pt}
    \acro{ASCI}{adaptive selected configuration interaction}
    \acro{CASCI}{complete active space configuration interaction}
    \acro{CI}{configuration interaction}
    \acro{GF(2)}{finite field of two elements}
    \acro{MPS}{matrix product state}
    \acro{PUI}{partial unary iteration}
    \acro{QDK}{Quantum Development Kit}
    \acro{QPE}{quantum phase estimation}
    \acro{QROM}{quantum read-only memory}
    \acro{REF}{row echelon form}
    \acro{SCI}[sCI]{selected configuration interaction}
\end{acronym}

\section{Introduction}
\label{sec:introduction}

Solving quantum many-body problems is widely regarded as one of the most promising applications of quantum computing~\cite{Lanyon2010, Hoefler2023}.
Algorithms such as \ac{QPE}~\cite{Kitaev1995, Chuang2010} provide a route to estimating electronic energies that are difficult to obtain classically in strongly correlated regimes, but their success depends on preparing an initial state with substantial overlap with the desired eigenstate~\cite{Chuang2010}.
Efficient quantum state preparation is therefore an essential primitive for such algorithms to achieve practical quantum utility.

State-preparation methods based on dissipative dynamics~\cite{Li2025, Ding2025}, adiabatic evolution~\cite{Guzik2005}, and eigenvalue filtering~\cite{Lin2020, Wang2023} can directly approximate target states on quantum hardware. However, because these methods rely on real-time quantum dynamics or Hamiltonian block-encodings, their circuit depth and ancilla overheads scale with the spectral gap and target accuracy, a cost borne primarily on the quantum device.
An alternative is to first compute an approximate state classically and then synthesize a dedicated preparation circuit, thereby moving the bulk of the work from quantum processor to classical precomputation.

This route is attractive because chemically relevant wavefunctions are sparse: only a small fraction of the configurations in the full Hilbert space carry significant weight, and configuration-based methods such as \ac{SCI}~\cite{Tubman2018} exploit this by identifying a compact set of dominant configurations that captures the essential correlation.
The target state (formally introduced in Section~\ref{sec:target-state}) is thus a superposition of $d \ll 2^n$ configurations (such as Slater determinants) on an $n$-qubit register, given as amplitude--bitstring pairs $\{(\vec{b}_j, \alpha_j)\}$. Our goal is to synthesize a circuit to prepare this state exactly while minimizing the circuit's non-Clifford gate count and ancillary qubit count, the resources most constrained on early fault-tolerant devices.

For direct circuit synthesis from such classical descriptions, general-purpose approaches include arbitrary isometry decomposition~\cite{Christandl2016}, the Grover--Rudolph method~\cite{Grover2002}, and standard \ac{QROM}-based methods~\cite{Berry2019, Low2024, Babbush2018}; for a general target state these must process up to $2^n$ amplitudes and therefore cost exponentially in $n$.
Methods tailored to the sparsity of the target state instead compress the $d$ nonzero-amplitude basis states onto a smaller \emph{dense register}, most often using permutations~\cite{Malvetti2021, Niels2021, Rupprecht2026, Ramacciotti2024}.
Malvetti et al.~\cite{Malvetti2021} permute the $d$ occupied basis states into a dense block on $\lceil\log_2 d\rceil$ qubits, though constructing the permutation requires heuristic searches over register partitions and mapping between representations requires many $\Toffoli$ gates; Rupprecht and W\"olk~\cite{Rupprecht2026} reduce this $\Toffoli$ cost using \ac{PUI}, which reuses control computations across consecutive basis states.
Vilmart et al.~\cite{Vilmart2026} recently published a method using Gauss--Jordan elimination on the binary matrix of basis states and synthesizing a $W$ state in the dense register.
For quantum-chemistry states, Fomichev et al.~\cite{Fomichev2024} uses \ac{QROM} and short $\operatorname{GF}(2)$-linear identifiers; Li and Luo~\cite{LiLuo2025} develop parallel sparse-state constructions with ancilla-dependent space-time tradeoffs.
Other approaches exploit target-state structure using decision diagrams~\cite{Mozafari2022}, state and angle trees~\cite{Araujo2023}, or Givens rotations for particle-number-conserving excitations~\cite{Greene-Diniz2025}, while the double-sparse method additionally exploits low Hamming weight within individual configurations~\cite{deVeras2022}.

In this work, we introduce a sparse state preparation method that exploits affine relationships among the configuration bitstrings over the \ac{GF(2)} to reduce the non-Clifford gate count and the ancillary qubit count.
Rather than relying on permutations, our method assembles the $d$ bitstrings as the columns of an $n \times d$ configuration matrix (formally introduced in Section~\ref{sec:config-matrix}) and reduces it with invertible affine transformations over \ac{GF(2)}: linear row additions together with translations by the all-ones vector. These operations compile directly to Clifford gates ($\CNOT$ and $\X$) and reduce the dense register from $n$ qubits to the rank $r$.
An optional binary encoding stage then uses non-Clifford conditional operations to compress the $d$ configurations onto the information-theoretic minimum of $\lceil\log_2 d\rceil$ qubits.
This Clifford-only affine reduction, carried out by Gaussian elimination together with all-ones-row removal, extracts the available structure, whereas binary encoding optionally trades $\Toffoli$ gates and ancillary qubits for a smaller dense state preparation cost when $r > \lceil\log_2 d\rceil$.

The workflow is modular: a classical stage records the compression gates, and the circuit prepares the compressed state with any dense state preparation subroutine before applying the recorded gates in reverse to expand it to the full register.
This gives rise to two variants, \emph{without binary encoding} (Gaussian elimination only) and \emph{with binary encoding} (both stages), which trade expansion-circuit resources against dense state preparation cost.

The remainder of this paper is organized as follows.
Section~\ref{sec:preliminaries} defines the mathematical objects and notation, Section~\ref{sec:methods} details the algorithm and derives the resource cost of each step, Section~\ref{sec:resource-analysis} compares these costs with prior methods, Section~\ref{sec:results} presents numerical benchmarks, and Section~\ref{sec:conclusions} discusses implications and limitations.
An open-source implementation is available through the \ac{QDK} chemistry application library~\cite{Baker2026}.

\section{Preliminaries}
\label{sec:preliminaries}

\subsection{Target state}
\label{sec:target-state}
Formally, each configuration bitstring introduced in Section~\ref{sec:introduction} is a distinct binary column vector $\vec{b}_j\in\operatorname{GF}(2)^n$, and the nonzero amplitudes $\alpha_j\in\mathbb{C}$ satisfy $\sum_{j=1}^{d}|\alpha_j|^2=1$.
The target state is
\begin{equation}\label{eq:target-state-v2}
    \ket{\psi} = \sum_{j=1}^{d} \alpha_j \ket{\vec{b}_j}.
\end{equation}
For a chemistry wavefunction under the Jordan--Wigner mapping~\cite{Wigner1928}, the $i$-th bit of $\vec{b}_j$ is $1$ if and only if orbital $i$ is occupied.
The algorithm introduced in this paper is also compatible with other fermion-to-qubit mappings, e.g., the Bravyi--Kitaev transformation~\cite{Bravyi-Kitaev2002,Love2012}; in such cases, $\ket{\psi}$ represents the encoded wavefunction in the computational basis of an $n$-qubit register.

\subsection{Configuration matrix}
\label{sec:config-matrix}
The configuration matrix $\mat{M} \in \operatorname{GF}(2)^{n \times d}$ is formed by assembling the $d$ bitstrings $\vec{b}_1, \dots, \vec{b}_d$ with nonzero amplitudes in the target state as columns:
\begin{equation}\label{eq:config-matrix}
    \mat{M} \;=\; \bigl[\,\vec{b}_1 \;\; \vec{b}_2 \;\; \cdots \;\; \vec{b}_d\,\bigr]
    \;\in\; \operatorname{GF}(2)^{n \times d},
\end{equation}
Thus, row $\mat{M}_{i,:}\in\operatorname{GF}(2)^d$ corresponds to qubit $i$, whereas column $\mat{M}_{:,j}=\vec{b}_j\in\operatorname{GF}(2)^n$ is configuration $j$.
Let $\mat{M}_{\mathrm{ref}}$ denote the \ac{REF} obtained after preprocessing and Gaussian elimination, as detailed in Section~\ref{sec:methods-v2-ge}. The rank $r$ is equal to the number of nonzero rows in $\mat{M}_{\text{ref}}$, with the following bounds:
\begin{equation}\label{eq:rank}
    \lceil\log_2 d\rceil\le r\le\min(n,d),
\end{equation}
The lower bound follows because preprocessing and Gaussian elimination preserve the $d$ distinct columns and an $r$-dimensional binary space contains at most $2^r$ vectors; the upper bound follows from the dimensions of $\mat{M}_{\mathrm{ref}}$.
The regime of interest is therefore $d\ll 2^n$ (sparsity) and $r\ll n$ (low rank).

Let $k$ denote the number of qubits with nonzero entries in the compressed configurations: $k=r$ for the variant \emph{without binary encoding} and $k=\lceil\log_2 d\rceil$ for the variant \emph{with binary encoding}.
These $k$ qubits form the \emph{dense register}; the remaining $n-k$ data qubits form the \emph{sparse register} and are fixed to $\ket{0}^{\otimes(n-k)}$ after compression.

\subsection{Matrix operations and quantum gates}
\label{sec:matrix-gate-operations}
The classical compression modifies $\mat{M}$ using row additions, affine row translations, conditional row operations, and column permutations.
Each row addition or affine row translation has a corresponding Clifford gate:
\begin{equation}\label{eq:row-gate-correspondence}
    \underbrace{\mat{M}_{q,:} \;\leftarrow\; \mat{M}_{q,:} + \mat{M}_{p,:}}_{\text{row addition over }\operatorname{GF}(2)}
    \;\longleftrightarrow\;
    \CNOT_{p\to q},
    \qquad
    \underbrace{\mat{M}_{p,:} \;\leftarrow\; \mat{M}_{p,:} + \mathbf{1}}_{\text{affine row translation}}
    \;\longleftrightarrow\;
    \X_p.
\end{equation}
Here, $\mathbf{1}\in\operatorname{GF}(2)^d$ is the all-ones row, and all additions are componentwise over $\operatorname{GF}(2)$.
Acting on the configurations (the columns of $\mat{M}$), these operations are invertible affine transformations over $\operatorname{GF}(2)$: the row additions are $\operatorname{GF}(2)$-linear, whereas a translation by $\mathbf{1}$ flips a qubit across every configuration and is affine rather than linear.
The Clifford compression therefore reduces $\mat{M}$ within the affine group over $\operatorname{GF}(2)$.

The binary-encoding variant also uses conditional row operations to reduce the number of nonzero rows.
Let $\vec{b}_{\mathrm{dense},j}$ denote column $j$ of rows corresponding to the dense register, let $\vec{\ell}$ be a dense-register pattern, and let $t$ index a row corresponding to a sparse register.
Such an operation updates
\begin{equation}\label{eq:conditional-row-operation}
    \mat{M}_{t,j} \;\leftarrow\; \mat{M}_{t,j}
    + \mathbf{1}\!\left[\vec{b}_{\mathrm{dense},j}=\vec{\ell}\right]
    \qquad \text{for each column }j,
\end{equation}
where $\mathbf{1}[\cdot]$ is the indicator function and addition is over $\operatorname{GF}(2)$.
The operation flips the entry in row $t$ exactly for columns whose dense pattern equals $\vec{\ell}$.
The corresponding quantum operation applies an $\lceil \log_2 d \rceil$-controlled $\X$ gate to target qubit $t$, controlled on the dense register being in the computational-basis state $\ket{\vec{\ell}}$.
To reduce the $\Toffoli$ gate count in compiling this multi-controlled gate, the binary-encoding procedure in Section~\ref{sec:methods-v2-be} exploits the column structure when implementing these conditional row operations.

A column permutation only reorders the classical pairs $(\vec{b}_j,\alpha_j)$ and therefore has no corresponding quantum gate.

\section{Algorithm details}
\label{sec:methods}

\subsection{Overview}
\label{sec:algorithm-overview}

Our method synthesizes the quantum circuit in two stages: classical matrix compression and quantum circuit construction.

\paragraph{Stage~1: Classical matrix compression}
The compression reduces the number of nonzero rows in $\mat{M}$ and records the corresponding quantum operations in two steps, applied sequentially:

\textbf{Gaussian elimination}
Using row operations defined in Section~\ref{sec:matrix-gate-operations}, Gaussian elimination reduces $\mat{M}$ to \ac{REF}:
\begin{equation}\label{eq:m-ref}
\mat{M}_{\text{ref}} = \begin{bmatrix} \mat{M}_{\text{GE}} \\ 0_{(n-r) \times d} \end{bmatrix},
\end{equation}
where $\mat{M}_{\text{GE}} \in \operatorname{GF}(2)^{r \times d}$ contains the $r$ nonzero rows and $0_{(n-r) \times d}$ is an all-zero block.
For the variant without binary encoding, this stage completes compression and sets the dense-register size to $k=r$.
See Section~\ref{sec:methods-v2-ge} for details.

\textbf{Binary encoding (optional)}
Binary encoding uses operations defined in Section~\ref{sec:matrix-gate-operations} to further compress $\mat{M}_{\text{ref}}$ from $r$ nonzero rows to $\lceil\log_2 d\rceil$ nonzero rows.
See Section~\ref{sec:methods-v2-be} for details.

Let $\mat{M}_{\text{dense}}\in\operatorname{GF}(2)^{k\times d}$ denote the nonzero block produced by the selected compression variant, and let its columns be $\vec{b}_{\text{dense},j}$.
The full sequence of quantum operations implements a global unitary transformation $U_{\text{compress}}$:
\begin{equation}\label{eq:gate-list}
\begin{aligned}
    U_{\text{compress}}\ket{\psi} &= U_{\text{compress}} \sum_{j=1}^{d} \alpha_j \ket{\vec{b}_j}
    = \sum_{j=1}^{d} \alpha_j \ket{\vec{b}_{\text{dense},j}} \ket{0}_S
    = \ket{\psi_{\text{dense}}}\ket{0}_S.
\end{aligned}
\end{equation}
Here, $\ket{\psi_{\text{dense}}}=\sum_j\alpha_j\ket{\vec{b}_{\text{dense},j}}$ is the compressed state on the dense register, and the sparse register (subscript $S$) is left in $\ket{0}^{\otimes(n-k)}$.

\paragraph{Stage~2: Quantum circuit construction}
The quantum circuit is constructed from the dense state $\ket{\psi_{\text{dense}}}$ and the sequence of quantum gates obtained from the classical matrix compression in two steps.

\textbf{Dense state preparation}
A unitary transformation $U_{\mathrm{dense}}$ prepares the compressed state $\ket{\psi_{\text{dense}}} = \sum_{j} \alpha_j \ket{\vec{b}_{\text{dense},j}}$ on $k$ qubits.
The choice of dense state preparation subroutine, and hence its gate count, is flexible: isometry decomposition~\cite{Christandl2016}, Grover--Rudolph~\cite{Grover2002}, or \ac{QROM}-based methods~\cite{Berry2019,Low2024,Babbush2018} can each be substituted depending on hardware constraints.

\textbf{Expansion}
The desired state $\ket{\psi}$ can be recovered from the dense state $\ket{\psi_{\text{dense}}}$ by applying $U_{\text{compress}}^\dagger$.
Thus, the quantum expansion circuit is constructed by applying the quantum operations corresponding to the recorded row operations in reverse order and replacing each operation with its adjoint.

The full state preparation unitary transformation is
\begin{equation}\label{eq:full-unitary}
    U \;=\; U_{\text{compress}}^\dagger \,\bigl(\, U_{\mathrm{dense}} \otimes I_{2^{n-k}} \,\bigr),
\end{equation}
where $I_{2^{n-k}}$ denotes the identity transformation on the $n-k$ sparse qubits.
Reading Eq.~\eqref{eq:full-unitary} from right to left, the quantum circuit first prepares $\ket{\psi_{\text{dense}}}$ on the $k$ dense qubits and then applies the expansion circuit on the full $n$-qubit register, recovering $\ket{\psi}$.

\subsection{Classical matrix compression}
\label{sec:methods-v2-compression}

This section describes the classical compression procedure in detail and derives the expansion-circuit resource cost associated with each step.

We illustrate the matrix compression workflow on a truncated \ac{SCI} wavefunction for the fluorine molecule (F\textsubscript{2}) using the top $d=8$ configurations and their coefficients from an \ac{SCI} calculation over $n=16$ qubits.

\begin{equation}
\label{eq:f2-wfn}
\begin{aligned}
    \ket{\psi} =\;&
        -0.6927\ket{1111100011111000}
        +0.4421\ket{1110101011101010}\\
        &+0.4192\ket{1101110011011100}
        -0.2785\ket{1100111011001110}\\
        &-0.2025\ket{1101101011101100}
        +0.1360\ket{1110110011011010}\\
        &+0.1058\ket{1111000111110001}
        -0.0261\ket{1101110011101100}
\end{aligned}
\end{equation}

In Eq.~\eqref{eq:f2-wfn}, the qubits are ordered $\ket{q_0 q_1 \dots q_{15}}$ from left to right, corresponding to the rows top-to-bottom in the configuration matrix in Figures~\ref{fig:f2-gaussian-elim} and~\ref{fig:f2-binary-encoding}. The top $k$ rows in the configuration matrix correspond to the dense qubits, while the remaining zero rows correspond to the sparse qubits.
The working example has $n=16$ data-qubit rows, $d=8$ configuration columns, rank $r=6$ after Gaussian elimination, and $\lceil \log_2 8 \rceil = 3$.

\subsubsection{Gaussian elimination over \texorpdfstring{$\operatorname{GF}(2)$}{GF(2)}}
\label{sec:methods-v2-ge}

Gaussian elimination reduces the configuration matrix $\mat{M}$ to the \ac{REF} $\mat{M}_{\text{ref}}$ with $r$ nonzero rows and $n-r$ zero rows.

\paragraph{Step~1: Preprocessing}
Before the main elimination loop, two preprocessing steps simplify the matrix: duplicate-row zeroing via row addition and all-ones-row zeroing via affine row translation.
These steps prune the search space and avoid redundant gates in the resulting circuit.
If rows $\mat{M}_{p,:}$ and $\mat{M}_{q,:}$ are identical, then $\mat{M}_{q,:} \leftarrow \mat{M}_{q,:} + \mat{M}_{p,:}$ zeros row~$q$.
An all-ones row satisfies $\mat{M}_{p,:}=\mathbf{1}$, so $\mat{M}_{p,:} \leftarrow \mat{M}_{p,:} + \mathbf{1}$ zeros it out.
The corresponding quantum operations are recorded according to Eq.~\eqref{eq:row-gate-correspondence}.

\paragraph{Step~2: Gaussian elimination}
The algorithm next scans the columns from left to right.
For each column, the first row containing a $1$ in the current column that has not yet served as a pivot is designated as the pivot for that column.
Every other non-pivot row $q$ that also contains a $1$ in the pivot column is updated via
\begin{equation}\label{eq:gf2-row-op}
    \mat{M}_{q,:} \;\leftarrow\; \mat{M}_{q,:} + \mat{M}_{p,:},
\end{equation}
where $p$ indexes the pivot row.

\begin{figure}
    \centering
    \renewcommand{\arraystretch}{1.1}
    \setlength{\tabcolsep}{3pt}%
    \resizebox{\textwidth}{!}{%
    \scalebox{0.5}{
    \begin{tikzpicture}[
        >=stealth, 
        every node/.style={inner sep=4pt, font=\scriptsize},
        arr/.style={->, thick},
        lbl/.style={font=\scriptsize, align=center}, 
    ]
        
        \node (m0) at (0,0) {$\begin{array}{r|*{8}{c}}
q_0   & 1&1&1&1&1&1&1&1\\
q_1   & 1&1&1&1&1&1&1&1\\
q_2   & 1&1&0&0&0&1&1&0\\
q_3   & 1&0&1&0&1&0&1&1\\
q_4   & 1&1&1&1&1&1&0&1\\
q_5   & 0&0&1&1&0&1&0&1\\
q_6   & 0&1&0&1&1&0&0&0\\
q_7   & 0&0&0&0&0&0&1&0\\
q_8   & 1&1&1&1&1&1&1&1\\
q_9   & 1&1&1&1&1&1&1&1\\
q_{10}& 1&1&0&0&1&0&1&1\\
q_{11}& 1&0&1&0&0&1&1&0\\
q_{12}& 1&1&1&1&1&1&0&1\\
q_{13}& 0&0&1&1&1&0&0&1\\
q_{14}& 0&1&0&1&0&1&0&0\\
q_{15}& 0&0&0&0&0&0&1&0\\
        \end{array}$};

        \node (m1) at (6.2,0) {$\begin{array}{r|*{8}{c}}
q_0   & 1&1&1&1&1&1&1&1\\
q_2   & 1&1&0&0&0&1&1&0\\
q_3   & 1&0&1&0&1&0&1&1\\
q_4   & 1&1&1&1&1&1&0&1\\
q_5   & 0&0&1&1&0&1&0&1\\
q_6   & 0&1&0&1&1&0&0&0\\
q_7   & 0&0&0&0&0&0&1&0\\
q_{10}& 1&1&0&0&1&0&1&1\\
q_{11}& 1&0&1&0&0&1&1&0\\
q_{13}& 0&0&1&1&1&0&0&1\\
q_{14}& 0&1&0&1&0&1&0&0\\
        \end{array}$};

        \node (m2) at (0,-6.0) {$\begin{array}{r|*{8}{c}}
q_2   & 1&1&0&0&0&1&1&0\\
q_3   & 1&0&1&0&1&0&1&1\\
q_4   & 1&1&1&1&1&1&0&1\\
q_5   & 0&0&1&1&0&1&0&1\\
q_6   & 0&1&0&1&1&0&0&0\\
q_7   & 0&0&0&0&0&0&1&0\\
q_{10}& 1&1&0&0&1&0&1&1\\
q_{11}& 1&0&1&0&0&1&1&0\\
q_{13}& 0&0&1&1&1&0&0&1\\
q_{14}& 0&1&0&1&0&1&0&0\\
        \end{array}$};

        \node (m3) at (6.2,-6.0) {$\begin{array}{r|*{8}{c}}
q_2   & 1&0&0&1&0&0&0&0\\
q_3   & 0&1&0&1&0&1&0&0\\
q_4   & 0&0&1&1&0&1&0&0\\
q_5   & 0&0&0&0&1&1&0&0\\
q_7   & 0&0&0&0&0&0&1&0\\
q_{10}& 0&0&0&0&0&0&0&1\\
        \end{array}$};

        
        \draw[arr] (m0) -- node[lbl, above] {Duplicate\\row\\zeroing} (m1);
        \draw[arr] (m2) -- node[lbl, above] {GF(2)\\Gaussian\\elimination} (m3);

        \path (m1.south) -- ++(0, -1.15) coordinate (vmid); 
        \draw[arr, rounded corners=6pt] 
            (m1.south) -- (vmid) -- (vmid -| m2.north) 
            node[lbl, midway, above] {All-ones row zeroing}
            -- (m2.north);

    \end{tikzpicture}}%
    }
    \caption{%
        Gaussian elimination of the F\textsubscript{2} configuration matrix.
        The three transformations are duplicate-row zeroing, all-ones-row zeroing, and Gaussian elimination over $\operatorname{GF}(2)$.
        Duplicate-row zeroing first zeros $q_1$, $q_8$, and $q_9$, which duplicate $q_0$; $q_{12}$, which duplicates $q_4$; and $q_{15}$, which duplicates $q_7$.
        The remaining all-ones row $q_0$ is then zeroed, and Gaussian elimination produces $\mat{M}_{\text{F}_2, \text{ref}}$ with six nonzero rows.
    }
    \label{fig:f2-gaussian-elim}
\end{figure}
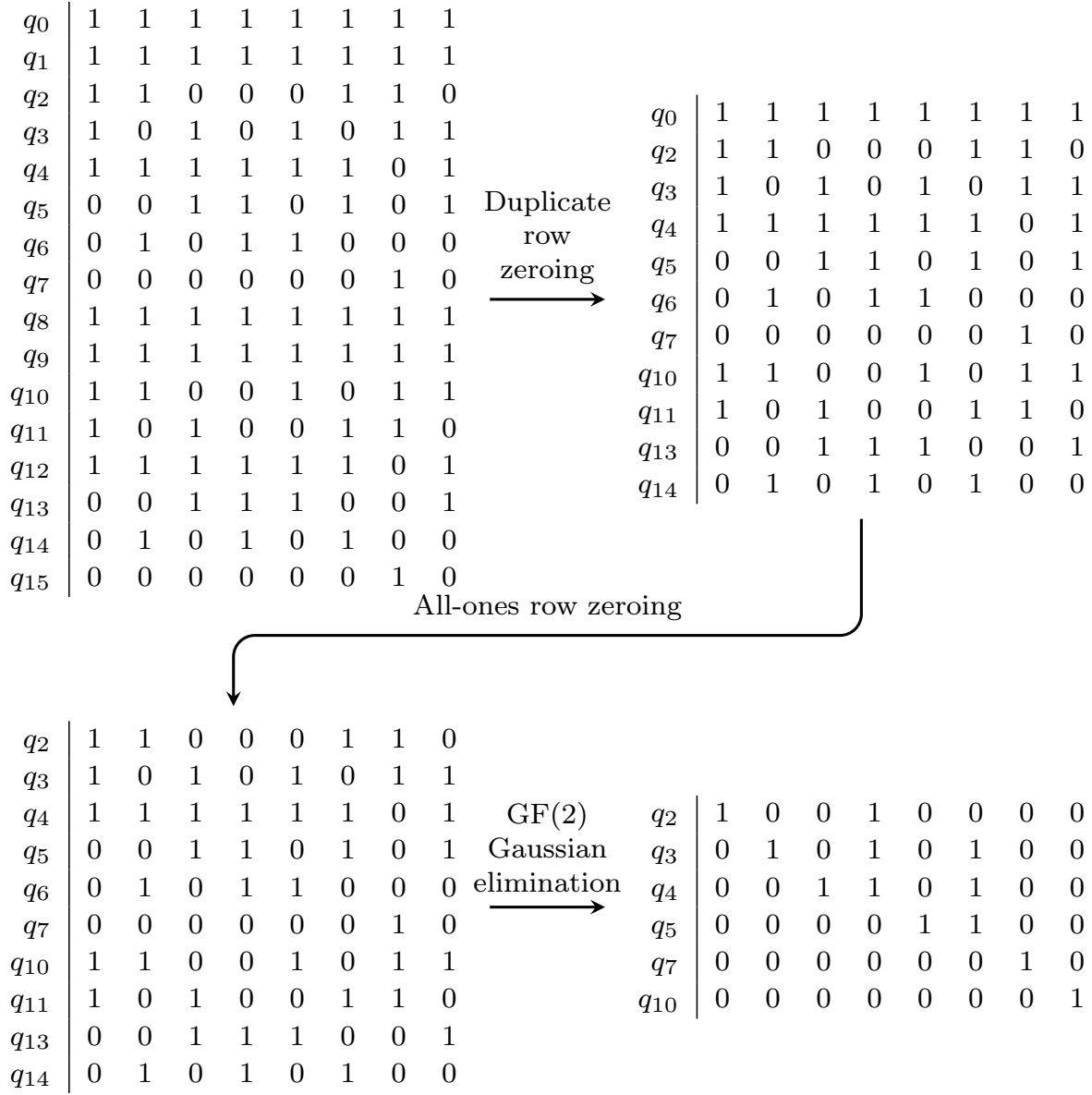

An illustration of this pass on the molecule F\textsubscript{2} is provided in Figure~\ref{fig:f2-gaussian-elim}.

\paragraph{Resource estimate}
For each of at most $r$ pivots, classical elimination scans at most $n$ rows of length $d$, requiring $\mathcal{O}(nrd)$ scalar field operations over $\operatorname{GF}(2)$.
At most $n-1$ row additions are recorded per pivot, in addition to at most $n$ initial $\X$ gates, so the corresponding expansion circuit contains $\mathcal{O}(nr)$ Clifford gates.
There are no $\Toffoli$ gates or ancillary qubits, and the dense state preparation acts on $r$ qubits.

\subsubsection{Binary encoding for diagonal and non-diagonal columns} 
\label{sec:methods-v2-be}

When $r > \lceil \log_2 d \rceil$, binary encoding further compresses $\mat{M}_{\text{ref}}$ from $r$ nonzero rows to $\lceil \log_2 d \rceil$ nonzero rows (see Section~\ref{sec:algorithm-overview} for an overview).
After a column permutation, $r$ linearly independent columns form an identity block $\mat{M}_{\text{diagonal}}\in\operatorname{GF}(2)^{r\times r}$, and the remaining $d-r$ columns form $\mat{M}_{\text{non-diagonal}}\in\operatorname{GF}(2)^{r\times(d-r)}$.
The two encoding steps apply the conditional row operations defined in Eq.~\eqref{eq:conditional-row-operation} efficiently by transforming these blocks into a binary-counter pattern and by exploiting the structure of their respective column blocks.

In the binary-counter pattern, $\vec{b}_{\mathrm{dense},j}$ is the fixed-width $k$-bit representation of $j-1$, with the most significant bit in the first dense row and the least significant bit in the last dense row, while every sparse-row entry is zero.
The dense columns therefore represent the consecutive integers $0,\dots,d-1$.
This ordering reduces the $\Toffoli$ gate count by allowing neighboring control states to reuse conditional prefixes.

\begin{figure}
    \centering
    \renewcommand{\arraystretch}{1.1}
    \setlength{\tabcolsep}{3pt}%
    \resizebox{\textwidth}{!}{%
    \scalebox{0.5}{
    \begin{tikzpicture}[
        >=stealth,
        every node/.style={inner sep=4pt, font=\scriptsize},
        arr/.style={->, thick},
        lbl/.style={font=\scriptsize, align=center},
    ]

        \node (fc0) at (0,0) {$\begin{array}{r|*{8}{c}}
q_2   & 1&0&0&1&0&0&0&0\\
q_3   & 0&1&0&1&0&1&0&0\\
q_4   & 0&0&1&1&0&1&0&0\\
q_5   & 0&0&0&0&1&1&0&0\\
q_7   & 0&0&0&0&0&0&1&0\\
q_{10}& 0&0&0&0&0&0&0&1\\
        \end{array}$};

        \node (fc_perm) at (6.8,0) {$\begin{array}{r|*{6}{c}|*{2}{c}}
q_2   & 1&0&0&0&0&0&1&0\\
q_3   & 0&1&0&0&0&0&1&1\\
q_4   & 0&0&1&0&0&0&1&1\\ 
q_5   & 0&0&0&1&0&0&0&1\\
q_7   & 0&0&0&0&1&0&0&0\\
q_{10}& 0&0&0&0&0&1&0&0\\
        \end{array}$};


        \node (fc1) at (0,-4) {$\begin{array}{r|*{6}{c}|*{2}{c}}
q_2   & 0&0&0&0&1&1&0&0\\
q_3   & 0&0&1&1&0&0&1&0\\
q_4   & 0&1&0&1&0&1&1&0\\ 
q_5   & 0&0&0&0&0&0&1&1\\
        \end{array}$};

        \node (fc2) at (6.8,-4) {$\begin{array}{r|*{6}{c}*{2}{c}}
q_2   & 0&0&0&0&1&1&1&1\\
q_3   & 0&0&1&1&0&0&1&1\\
q_4   & 0&1&0&1&0&1&0&1\\
        \end{array}$};


        \draw[arr] (fc0.east) -- node[lbl, above] {Column\\permutation} (fc_perm.west);

        \path (fc_perm.south) -- ++(0, -1.0) coordinate (vmid);
        \draw[arr, rounded corners=6pt]
            (fc_perm.south) -- (vmid) -- (vmid -| fc1.north)
            node[lbl, midway, above] {Diagonal encoding}
            -- (fc1.north);

        \draw[arr] (fc1.east) -- node[lbl, above] {Non-diagonal\\column\\encoding} (fc2.west);

    \end{tikzpicture}}%
    }
    \caption{%
    Binary encoding of the F\textsubscript{2} \ac{REF} matrix, including the diagonal and non-diagonal encoding steps.
    Since $d=8$ configurations require only $\lceil \log_2 8 \rceil = 3$ rows for addressing, the top three rows are retained as the nonzero rows corresponding to the dense qubits.
    Columns $c_0, c_1, c_2, c_4, c_6$, and $c_7$ are first permuted to the left to form the diagonal block, while the non-diagonal columns $c_3$ and $c_5$ are moved to the right.
    Diagonal encoding (Step~1) transforms the identity block into binary-counter form on the nonzero rows by assigning contiguous binary labels $0$--$5$ while zeroing the remaining rows. Note that this step also modifies the entries of the non-diagonal columns, so the matrix is updated accordingly.
    Non-diagonal encoding (Step~2) encodes $c_3$ and $c_5$ using the same binary-counter pattern and clears the remaining nonzero entries through a batch-zeroing operation, yielding the final compressed matrix.
    }
    \label{fig:f2-binary-encoding}
\end{figure}
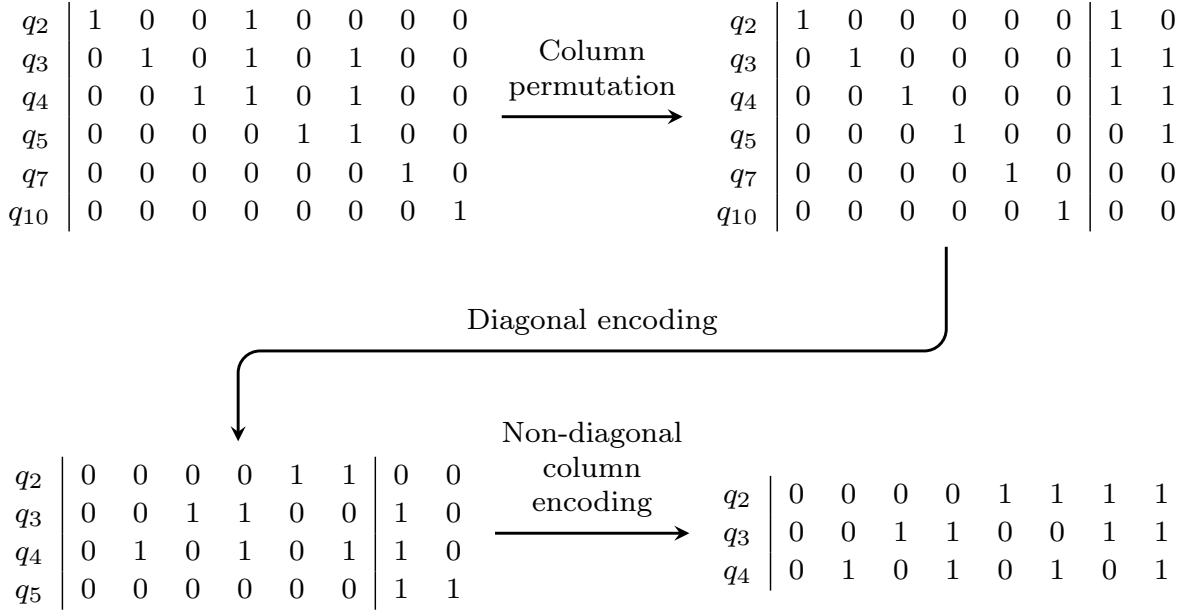

Diagonal encoding (Step~1) applies a specialized recursive routine to $\mat{M}_{\text{diagonal}}$, assigning labels $0, \dots, r-1$ while zeroing the sparse-row entries of that block.
Non-diagonal encoding (Step~2) assigns labels $r, \dots, d-1$ to $\mat{M}_{\text{non-diagonal}}$ using row additions, then applies \ac{PUI} to zero the sparse-row entries of that block in batches.
Figure~\ref{fig:f2-binary-encoding} illustrates both steps for the F\textsubscript{2} configuration matrix.

\paragraph{Step~1: Diagonal encoding} 

First, a sequence of row additions transforms the identity block into a unary matrix (an upper-triangular matrix whose entries are all ones). These row operations correspond to a reverse $\CNOT$ ladder.
Second, a recursive procedure converts the unary pattern into a binary-counter pattern. 

The recursive procedure pairs adjacent rows and, for each pair, uses two row additions Eq.~\eqref{eq:row-gate-correspondence} to accumulate the pair's parity into a designated accumulator row (the lowest-indexed available row), followed by a conditional row operation in Eq.~\eqref{eq:conditional-row-operation} that zeros out the even-indexed row of the pair.
The corresponding circuit implementation for each pair consists of two $\CNOT$ gates and one $\Toffoli$ gate controlled on the accumulator qubit and the odd-indexed qubit, targeting the even-indexed qubit.
The odd-indexed rows are retained for the next iteration.
If the number of rows being processed is odd, the final row is accumulated and carried to the next iteration.
At each level of the recursion, the number of processing rows is roughly halved, while the accumulator rows are incrementally updated.

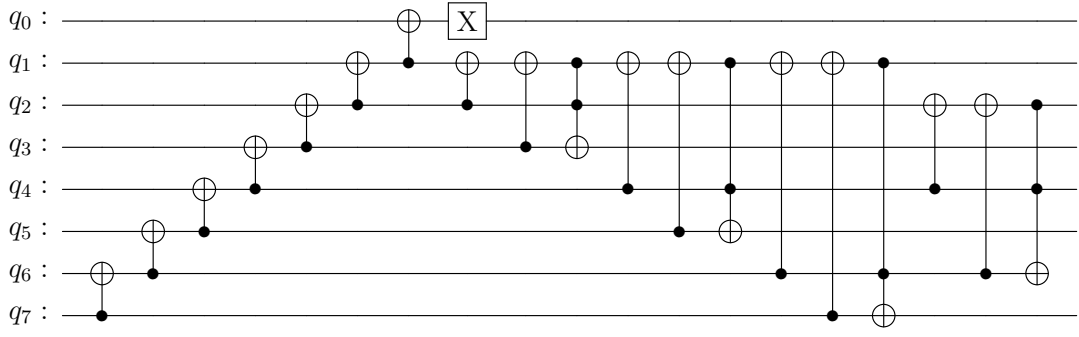
\begin{figure}
\centering
\scalebox{0.9}{
\Qcircuit @C=1.0em @R=0.2em @!R { \\
	 	\nghost{{q}_{0} :  } & \lstick{{q}_{0} :  } & \qw & \qw & \qw & \qw & \qw & \qw & \targ & \gate{\mathrm{X}} & \qw & \qw & \qw & \qw & \qw & \qw & \qw & \qw & \qw & \qw & \qw & \qw \\
	 	\nghost{{q}_{1} :  } & \lstick{{q}_{1} :  } & \qw & \qw & \qw & \qw & \qw & \targ & \ctrl{-1} & \targ & \targ & \ctrl{1} & \targ & \targ & \ctrl{3} & \targ & \targ & \ctrl{5} & \qw & \qw & \qw & \qw\\
	 	\nghost{{q}_{2} :  } & \lstick{{q}_{2} :  } & \qw & \qw & \qw & \qw & \targ & \ctrl{-1} & \qw & \ctrl{-1} & \qw & \ctrl{1} & \qw & \qw & \qw & \qw & \qw & \qw & \targ & \targ & \ctrl{2} & \qw \\
	 	\nghost{{q}_{3} :  } & \lstick{{q}_{3} :  } & \qw & \qw & \qw & \targ & \ctrl{-1} & \qw & \qw & \qw & \ctrl{-2} & \targ & \qw & \qw & \qw & \qw & \qw & \qw & \qw & \qw & \qw & \qw \\
	 	\nghost{{q}_{4} :  } & \lstick{{q}_{4} :  } & \qw & \qw & \targ & \ctrl{-1} & \qw & \qw & \qw & \qw & \qw & \qw & \ctrl{-3} & \qw & \ctrl{1} & \qw & \qw & \qw & \ctrl{-2} & \qw & \ctrl{2} & \qw \\
	 	\nghost{{q}_{5} :  } & \lstick{{q}_{5} :  } & \qw & \targ & \ctrl{-1} & \qw & \qw & \qw & \qw & \qw & \qw & \qw & \qw & \ctrl{-4} & \targ & \qw & \qw & \qw & \qw & \qw & \qw & \qw \\
	 	\nghost{{q}_{6} :  } & \lstick{{q}_{6} :  } & \targ & \ctrl{-1} & \qw & \qw & \qw & \qw & \qw & \qw & \qw & \qw & \qw & \qw & \qw & \ctrl{-5} & \qw & \ctrl{1} & \qw  & \ctrl{-4} & \targ \\
	 	\nghost{{q}_{7} :  } & \lstick{{q}_{7} :  } & \ctrl{-1} & \qw & \qw & \qw & \qw & \qw & \qw & \qw & \qw & \qw & \qw & \qw & \qw & \qw & \ctrl{-6} & \targ & \qw & \qw & \qw & \qw\\
\\ }}
    \caption{%
        Diagonal encoding circuit corresponding to the matrix operation that compresses an identity block $I = [\delta_{ij}]$ for $1 \le i, j \le 8$ into a binary-counter pattern in the nonzero rows associated with $q_1, q_2, q_4$, while zeroing out the rows associated with $q_0, q_3, q_5, q_6, q_7$. 
        A reverse $\CNOT$ ladder and an $\X$ gate first transform the identity block into unary form. 
        The circuit then recursively aggregates the qubits associated with row pairs into an accumulator and zeros out the even-indexed rows using $\CNOT$ and $\Toffoli$ gates.
        At the end, the qubits $q_4, q_2, q_1$ (most to least significant) hold the binary counter $0, 1, \dots, 7$, while the remaining qubits are in the $\ket{0}$ state.
    }
    \label{fig:diagonal-encoding}
\end{figure}

At the end, in the first $r$ columns, the $\lceil \log_2 r \rceil$ dense rows carry contiguous binary-counter pattern $0, 1, \dots, r-1$, and the remaining rows are zero.
See Figure~\ref{fig:diagonal-encoding} for a standalone diagonal encoding circuit on an $r=8$ identity block, which illustrates the same procedure as Step~1 in Figure~\ref{fig:f2-binary-encoding} at a larger size than the $r=6$ F\textsubscript{2} block.

\paragraph{Step~2: Non-diagonal encoding} 
The non-diagonal encoding assigns binary-counter patterns \allowbreak $r, \dots, d-1$ to the $d-r$ non-diagonal columns and zeros all remaining sparse-row entries.
The algorithm follows the batched-pivoting method developed by Rupprecht and W\"olk~\cite{Rupprecht2026}.

After diagonal encoding, the $r$ diagonal columns carry binary labels $0, \dots, r-1$ in the dense rows, but the $d-r$ non-diagonal columns still have nonzero entries in the sparse rows.
Without batching, zeroing each non-diagonal column would require a $\lceil \log_2 d \rceil$-controlled $\X$ gate, totaling $d-r$ multi-controlled gates.
The \ac{PUI} subroutine reduces this cost by processing multiple columns simultaneously.

This algorithm proceeds column by column and accumulates the processed columns into batches of size~$m$.
For each non-diagonal column, a sparse row containing a $1$ in that column is selected as a pivot, analogous to choosing a pivot row in Gaussian elimination.
Row additions then eliminate all other entries $1$ in the sparse portion of the column and assign the next contiguous binary label in the dense portion.
The processed column is then added to the current batch.

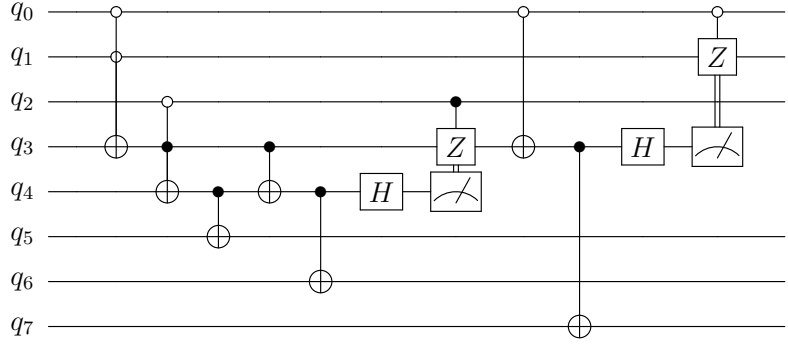
\begin{figure}
    \centering
    \scalebox{0.9}{
    \Qcircuit @C=1.0em @R=0.2em @!R { \\
        \lstick{q_0} & \qw & \ctrlo{3} & \qw      & \qw      & \qw      & \qw      & \qw      & \qw      & \ctrlo{3} & \qw      & \qw      & \ctrlo{1} & \qw \\
        \lstick{q_1} & \qw & \ctrlo{2}& \qw      & \qw      & \qw      & \qw      & \qw      & \qw      & \qw      & \qw      & \qw      & \gate{Z} & \qw \\
        \lstick{q_2} & \qw & \qw      & \ctrlo{2}& \qw      & \qw      & \qw      & \qw      & \ctrl{1}& \qw      & \qw      & \qw      & \qw      & \qw \\
        \lstick{q_3} & \qw & \targ    & \ctrl{1} & \qw      & \ctrl{1} & \qw      & \qw      & \gate{Z} & \targ    & \ctrl{4} & \gate{H} & \meter \cwx[-2] & \push{\rule{1em}{0pt}} \\
        \lstick{q_4} & \qw & \qw      & \targ    & \ctrl{1} & \targ    & \ctrl{2} & \gate{H} & \meter \cwx[-1] & \push{\rule{1em}{0pt}}   & \push{\rule{1em}{0pt}}   & \push{\rule{1em}{0pt}}   & \push{\rule{1em}{0pt}}   & \push{\rule{1em}{0pt}} \\
        \lstick{q_5} & \qw & \qw      & \qw      & \targ    & \qw      & \qw      & \qw      & \qw      & \qw      & \qw      & \qw      & \qw      & \qw \\
        \lstick{q_6} & \qw & \qw      & \qw      & \qw      & \qw      & \targ    & \qw      & \qw      & \qw      & \qw      & \qw      & \qw      & \qw \\
        \lstick{q_7} & \qw & \qw      & \qw      & \qw      & \qw      & \qw      & \qw      & \qw      & \qw      & \targ    & \qw      & \qw      & \qw \\
    }}
    \caption{Partial unary iteration subroutine for the interval $[|0\rangle, |1\rangle, |2\rangle]$. The first 3 qubits ($q_0 \dots q_2$) act as the dense address qubits, the middle 2 qubits ($q_3, q_4$) are ancillary qubits, and the last 3 qubits ($q_5 \dots q_7$) are the sparse target qubits. The ancillary qubits are freed using measurement-based uncomputation~\cite{Gidney2018}. Note that the $|00\rangle$ prefix state of $q_0$ and $q_1$ cached in the ancilla $q_3$ is reused for both address states 0 and 1, and the state stored in $q_4$ is flipped by a $\CNOT$ from $q_3$ to target state 1. This demonstrates how the $\Toffoli$ gate count is reduced with batches of contiguous address states.}
    \label{fig:pui-interval-circuit}
\end{figure}

Once a batch reaches size $m$ or exhausts the available sparse rows, all sparse-row entries in the batch are zeroed simultaneously using a \ac{PUI} subroutine.
As defined in Section~\ref{sec:matrix-gate-operations}, the subroutine uses the dense qubits as address qubits and the sparse qubits as targets, zeroing each target entry conditioned on its binary-counter label.
Because each batch contains a contiguous interval of labels, \ac{PUI} reuses shared prefix conditions and reduces the $\Toffoli$ gate count relative to addressing each column independently.
The ancillary qubits are freed using measurement-based uncomputation~\cite{Gidney2018}.
See Figure~\ref{fig:pui-interval-circuit} for an example circuit.

\paragraph{Resource estimate}

In diagonal encoding, the identity-to-unary transformation uses $r-1$ $\CNOT$ gates and one $\X$ gate.
The subsequent recursive encoding uses one $\Toffoli$ gate and two $\CNOT$ gates per processed row pair.
The identity-to-unary transformation may repeat row additions performed during Gaussian elimination; reducing the pivot columns directly to unary form avoids these redundant $\CNOT$ gates. Compared with applying \ac{PUI} to the identity block, this routine exploits its fixed one-hot structure and requires only $r-\lceil\log_2 r\rceil-1$ $\Toffoli$ gates without additional workspace.

Let $s=d-r$ denote the number of non-diagonal columns.
For $s\ge 2$, non-diagonal encoding partitions the columns into batches of size $m$ and applies the batched isometry subroutine of Rupprecht and W\"olk~\cite{Rupprecht2026}, where the estimated $\Toffoli$-gate count is
\begin{equation}\label{eq:rupprecht-isometry-bound}
        \mathcal{B}(s,m)
        = \left\lceil\frac{s}{m}\right\rceil
            \left(2m+\frac{1}{2}\log_2\!\left(\frac{\tilde{s}}{m}\right)-3\right)
            +\log_2\!\left(\frac{\tilde{s}^{2}}{m}\right),
        \qquad
        \tilde{s}=2^{\lceil\log_2s\rceil}.
\end{equation}
Here, $m$ is a power-of-two batch size satisfying $1\le m\le\tilde{s}$.
Because the labels occupy the shifted interval $r,\dots,d-1$ within the $\lceil\log_2 d\rceil$-bit dense register, rather than a tight $\lceil\log_2 s\rceil$-bit register, $\mathcal{B}(s,m)$ is an estimate; Section~\ref{sec:results} reports exact compiled counts in numerical benchmarks.
The non-diagonal subroutine uses $\max\{0,\lceil\log_2s\rceil-1\}$ workspace qubits, of which up to $n-r$ are supplied by data qubits zeroed during Gaussian elimination.

Combining the two steps, binary encoding for $s\ge 2$ uses an estimated
\begin{equation}\label{eq:binary-encoding-resource-estimate}
    r-\lceil\log_2r\rceil-1+\mathcal{B}(s,m)
\end{equation}
$\Toffoli$ gates and requires $\max\{0,\lceil\log_2s\rceil-1-(n-r)\}$ additional ancillary qubits.
When $s=0$, the non-diagonal step contributes no gates or ancillary qubits; when $s=1$, one $\lceil\log_2 d\rceil$-controlled $\X$ gate replaces the $\mathcal{B}(s,m)$ term and requires no additional ancillary qubits.

\subsection{Resource analysis and comparison to prior methods}
\label{sec:resource-analysis}

We compare the expansion-circuit resource bounds of both variants with those of prior sparse state preparation methods for $d\geq2$ in Table~\ref{tab:resource-comparison}, excluding dense state preparation.
Here, we include only expansion methods that can be composed with an arbitrary dense state preparation routine, excluding Vilmart et al.~\cite{Vilmart2026}'s construction using a one-hot $W$-state preparation and Rupprecht and W\"olk~\cite{Rupprecht2026}'s co-optimized construction of permutation and the dense state preparation. We retain their method-agnostic batched-pivoting expansion as a baseline.
\begin{table}[ht]
\centering
\small
\caption{Expansion-circuit resource bounds for sparse state preparation. The binary-encoding estimate applies for $d-r\ge 2$. $\mathcal{B}(s,m)$ is defined in Eq.~\eqref{eq:rupprecht-isometry-bound}. The Ramacciotti et al. entry is the worst case of $d$ two-cycles obtained from their per-cycle gate count.}
\label{tab:resource-comparison}
\resizebox{\textwidth}{!}{%
\begin{tabular}{@{}lccc@{}}
\toprule
Method & $\Toffoli$-gate count & Ancillary qubits & Dense qubits \\
\midrule
Ours (w/o binary encoding) & $0$ & $0$ & $r$ \\
Ours (w/ binary encoding) & $r-\lceil\log_2r\rceil-1+\mathcal{B}(d-r,m)$ & \shortstack{$0$ if $d=r$; else\\$\max\{0,\lceil\log_2(d-r)\rceil-1-(n-r)\}$} & $\lceil\log_2 d\rceil$ \\
Rupprecht--W\"olk~\cite{Rupprecht2026} & $\mathcal{B}(d,m)$ & $\lceil\log_2d\rceil-1$ & $\lceil\log_2 d\rceil$ \\
Ramacciotti et al.~\cite{Ramacciotti2024} & $6d(n-1)$ & $n$ & $\lceil\log_2 d\rceil$ \\
\bottomrule
\end{tabular}}
\end{table}

Table~\ref{tab:resource-comparison} highlights two operating regimes for our method.
When $r$ is close to $\lceil\log_2 d\rceil$, Gaussian elimination alone is attractive because its expansion circuit uses no $\Toffoli$ gates or ancillary qubits, while its dense register is only moderately larger than the minimum size.
When $r$ substantially exceeds $\lceil\log_2 d\rceil$, binary encoding can reduce the dense state preparation cost by shrinking the dense register from $r$ to $\lceil\log_2 d\rceil$ qubits, at the expense of the $\Toffoli$ gates and workspace shown in the table.

Relative to Rupprecht--W\"olk, Gaussian elimination reduces the input to the batched isometry from $d$ columns to the $d-r$ non-diagonal columns.
For a fixed batch size $m$ and $d-r\ge 2$, our binary-encoding expansion uses fewer $\Toffoli$ gates whenever $r-\lceil\log_2r\rceil-1+\mathcal{B}(d-r,m)<\mathcal{B}(d,m)$.
It requires no additional ancillary qubits when the $n-r$ data qubits zeroed by Gaussian elimination provide the required workspace, namely when $n-r\geq\lceil\log_2(d-r)\rceil-1$ for $d>r$.
The Ramacciotti et al. bound instead retains a multiplicative dependence on the full register size $n$, giving it a higher $\Toffoli$ gate count and ancillary qubit count than our method.
These bounds exclude dense state preparation; the end-to-end counts in Section~\ref{sec:results} additionally include it, which is where the without-binary-encoding variant incurs its cost.

\section{Results}
\label{sec:results}

We benchmark our method in two variants, \emph{without binary encoding} and \emph{with binary encoding}, against two recent sparse state preparation methods with publicly available implementations: the batched-pivoting algorithm of Rupprecht--Wölk~\cite{Rupprecht2026} and the permuted Grover--Rudolph method of Ramacciotti et al.~\cite{Ramacciotti2024}, both implemented in \textsc{Qualtran}~\cite{harrigan2024qualtran}.
All methods use the same dense state preparation primitive in \ac{QDK}~\cite{Shende2006} to ensure a fair comparison.

To evaluate scaling, we generate random binary matrices representing the Jordan--Wigner encoding with half filling ($n_\mathrm{electrons} = n_\mathrm{orb} = n/2$) and the number of configurations set equal to the number of qubits.
We also include chemically derived instances, extracted from the \ac{QDK} Chemistry workflow, as marker overlays in the scaling plots \cite{Baker2026}.
Molecular geometries were optimized using ORCA~\cite{ORCA6}, and the active space was selected using the MACIS \ac{ASCI} solver~\cite{williams23_distributed_asci} and autoCAS refinement~\cite{Stein2019autoCAS}, followed by a final \ac{CASCI} calculation \cite{williams23_distributed_asci} with the cc-pVDZ basis set \cite{Dunning1989}.

\begin{figure}
    \centering
    \begin{adjustwidth}{-0.5in}{-0.5in}
    \includegraphics[width=\linewidth]{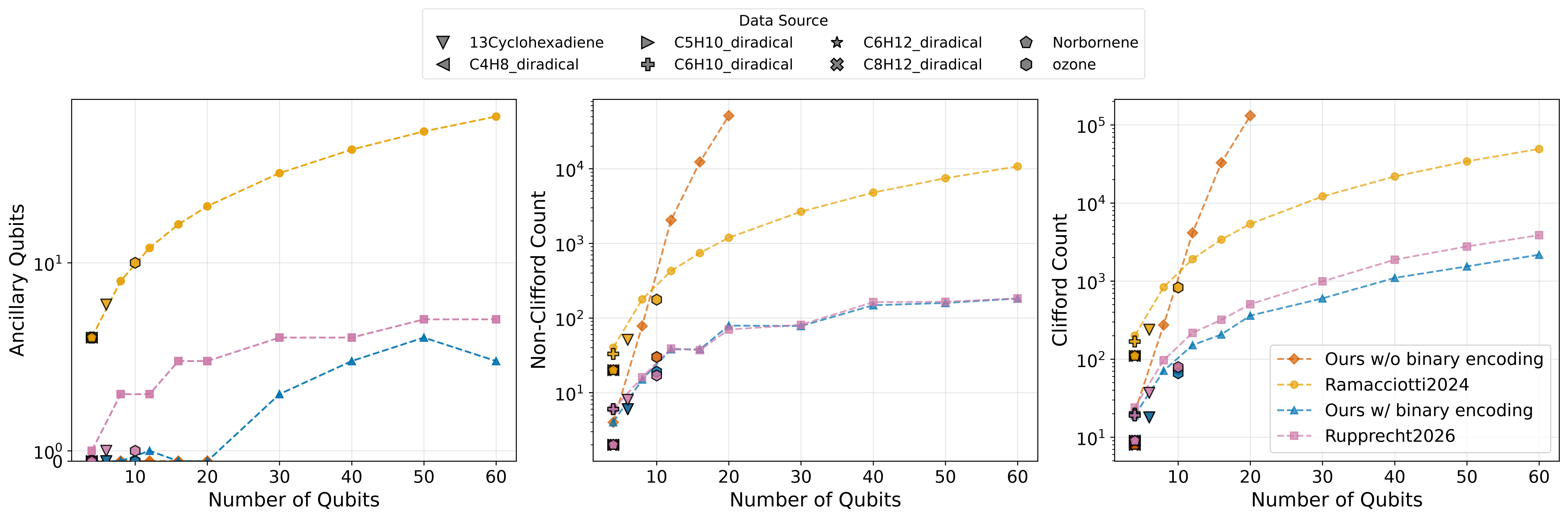}
    \end{adjustwidth}
    \caption{Scaling with qubit count for random Jordan--Wigner encoding matrices, with chemically derived instances overlaid as markers.
    Dashed lines denote the scaling baselines for random matrices.
    Overlaid individual points (with different marker shapes) represent the specific chemically derived molecular systems.
    Marker colors correspond to the algorithm used.
    (a) Ancillary qubit count. (b) Non-Clifford-gate count. (c) Clifford-gate count.
    }
    \label{fig:qubit-scaling}
\end{figure}

Figure~\ref{fig:qubit-scaling} reveals the trade-off between the two variants. Without binary encoding, the method is competitive only for very small systems: when the rank $r$ exceeds $\lceil \log_2 d \rceil$, the dense state preparation circuit acts on $r$ qubits rather than the minimum $\lceil \log_2 d \rceil$, incurring a significantly larger gate count.
By contrast, our method with binary encoding scales to larger instances with fewer ancillary qubits than both reference methods, and with non-Clifford gate counts comparable to those of Rupprecht--Wölk~\cite{Rupprecht2026}.
The non-Clifford gate count here includes the rotations in the dense state preparation circuit and the $\Toffoli$ gates in the expansion circuit.
The compilation of these non-Clifford gates depends on the specific error-correction code and qubit architecture~\cite{Gidney2024, Cody2013, Kliuchnikov2023}.
Our method with binary encoding also exhibits a moderately lower Clifford gate count than the Rupprecht--W\"olk method (Figure~\ref{fig:qubit-scaling}b).

On the sampled molecular systems, both variants perform at least as well as on random matrices: the gate and ancillary qubit counts required to prepare chemical wavefunctions are less than or equal to the random-matrix baselines.
The favorable performance on these chemical systems reflects the structure of the determinant spaces produced by ASCI, whose search generates candidate determinants through single and double excitations from dominant core configurations~\cite{williams23_distributed_asci}.
In our benchmarks, the resulting shared excitation patterns produce affine dependencies among the rows of the configuration matrix, enabling the Clifford compression to achieve a greater matrix reduction, which in turn reduces the computational effort for binary encoding or dense state preparation.
A similar physical structure has recently been exploited to lower quantum resource costs in \ac{MPS} preparation~\cite{Rupprecht2026MPS}.

\begin{figure}
    \centering
    \begin{adjustwidth}{-0.5in}{-0.5in}
    \includegraphics[width=\linewidth]{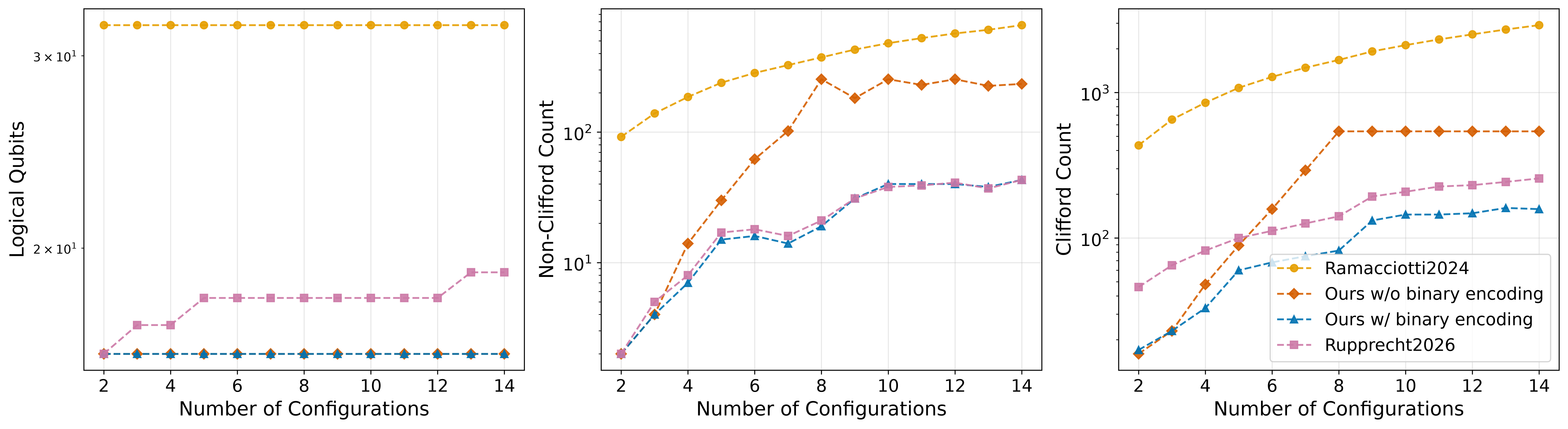}
    \end{adjustwidth}
    \caption{F\textsubscript{2} benchmark ($n=16$): gate and qubit scaling versus number of configurations. (a) Total logical qubit count (data + ancillary qubits). (b) Non-Clifford-gate count. (c) Clifford-gate count.}
    \label{fig:f2-gate-counts}
\end{figure}

Figure~\ref{fig:f2-gate-counts} illustrates the gate count and ancillary qubit scaling as a function of the number of configurations for a fixed number of qubits in F\textsubscript{2}.
For small $d \le 4$, both variants achieve the lowest non-Clifford gate counts and require no ancillary qubits.
As $d$ increases, our method without binary encoding incurs a steep dense state preparation penalty because the dense state preparation circuit acts on $r$ qubits, whereas the non-Clifford gate count for our method with binary encoding remains low, comparable to those of Rupprecht--Wölk and far below those of Ramacciotti et al.
Our method without binary encoding still requires no ancillary qubits, and our method with binary encoding requires fewer ancillary qubits than the other methods tested in this paper.

\begin{figure}
    \centering
    \includegraphics[width=0.7\textwidth]{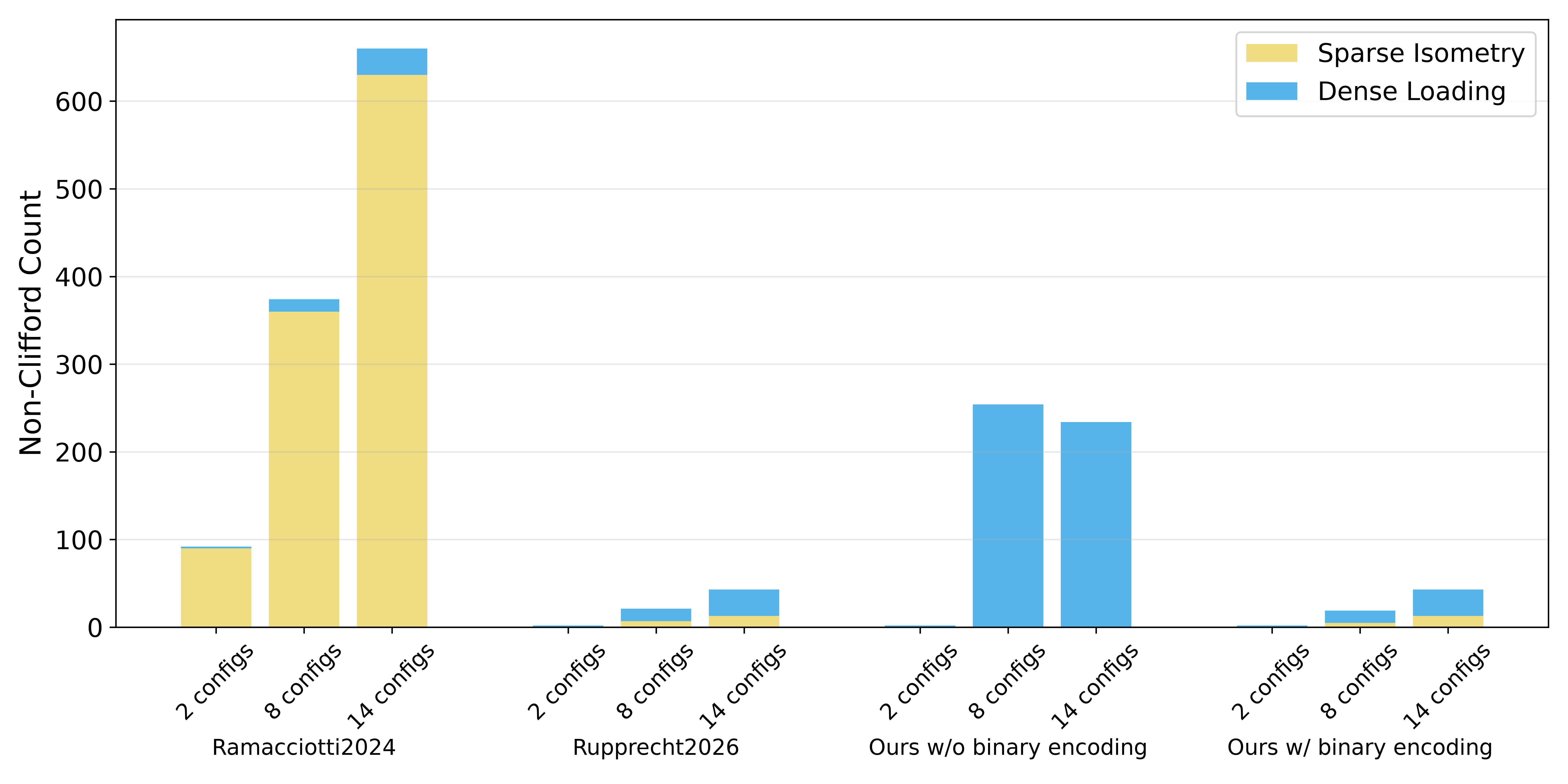}
    \caption{Breakdown of non-Clifford gates into expansion circuit (yellow) and dense state preparation circuit (blue) contributions for the F\textsubscript{2} molecule at $d=2,8,14$ configurations.}
    \label{fig:f2-stacked-gate-counts}
\end{figure}

Figure~\ref{fig:f2-stacked-gate-counts} decomposes the non-Clifford gate count into expansion circuit and dense state preparation circuit contributions for selected numbers of configurations ($d = 2, 8, 14$) in F\textsubscript{2}.
The decomposition highlights the central tradeoff: without binary encoding, our method achieves zero non-Clifford gates in the expansion circuit but incurs a large dense state preparation gate count at higher $d$; with binary encoding, it introduces additional $\Toffoli$ gates in the expansion circuit to substantially reduce the non-Clifford cost of dense state preparation.

\section{Conclusions}
\label{sec:conclusions}

In this work, we introduced a sparse state preparation method that improves on the compression by permutations used in prior work~\cite{Malvetti2021, Rupprecht2026, Ramacciotti2024} by exploiting affine relationships among configuration bitstrings.
The first stage uses Clifford-only Gaussian elimination to compress the configuration matrix to its $r$ nonzero rows, where $r$ is the rank.
The optional binary encoding stage then trades additional $\Toffoli$ gates and ancillary qubits for further compression.
It reduces the dense register to $\lceil\log_2 d\rceil$ qubits, the information-theoretic minimum.
Among the sparse state preparation methods evaluated here, the resulting circuits achieve the lowest ancillary qubit count.

The modular workflow enables a trade-off between ancillary qubit count and non-Clifford gate count: Gaussian elimination alone is efficient when $r\approx\lceil\log_2 d\rceil$, whereas binary encoding becomes advantageous when the reduction in dense state preparation cost outweighs its $\Toffoli$-gate and ancillary-qubit overhead.
As $r$ grows relative to $\lceil\log_2 d\rceil$, binary encoding provides a larger reduction in dense state preparation cost, although its ancillary-qubit and $\Toffoli$-gate-count advantage over Rupprecht--W\"olk~\cite{Rupprecht2026} narrows.

The efficiency of the method depends on two properties of the input: state sparsity ($d \ll 2^n$) and low rank ($r \ll n$).
The first condition is naturally satisfied by chemically relevant states produced by active-space methods such as \ac{SCI}~\cite{Tubman2018}.
Our numerical experiments indicate that the second condition often holds because the dominant configurations share significant structural patterns, frequently differing from a reference configuration by a few orbital excitations.
These shared patterns produce duplicate, all-ones, or linearly dependent rows, which the method exploits to obtain a Clifford-efficient compression circuit.
More broadly, the Gaussian elimination used for compression is connected to the row-reduction primitive underlying classical linear error-correcting codes~\cite{Reed1960, Bose1960, Gallager1962}.

Overall, our method provides a modular approach for preparing complex chemical states in early fault-tolerant quantum algorithms such as \ac{QPE}.
An open-source implementation is available through the \ac{QDK} Chemistry application toolkit~\cite{Baker2026}.

\section*{Acknowledgements}
We thank Aarthi Sundaram, Matthias Christandl, Matthias Troyer, Shuchen Zhu, and Wim van Dam for valuable comments on the manuscript.

\section*{Data Availability}
The data and scripts used to generate the figures presented in this study are available at https://github.com/microsoft/qdk-chemistry-data.
The core algorithmic implementation is accessible via the open-source \ac{QDK} Chemistry application library.
Appendix~\ref{sec:appendix-code} illustrates the end-to-end workflow for preparing a multi-configurational wavefunction on a quantum computer using \ac{QDK} Chemistry.

\bibliographystyle{unsrt}
\bibliography{references}

\appendix
\section{Code samples using QDK Chemistry}
\label{sec:appendix-code}

To run the following example, install \ac{QDK} Chemistry from source. 
\begin{verbatim}
git clone https://github.com/microsoft/qdk-chemistry.git
\end{verbatim}
Open the repository in VS Code and reopen in the VS Code Dev Container. See the \href{https://github.com/microsoft/qdk-chemistry/blob/main/INSTALL.md}{INSTALL.md} for details.

\begin{figure}[H]
\begin{lstlisting}[language=Python]
import numpy as np
from qdk_chemistry.algorithms import create
from qdk_chemistry.data import Element, Structure
from qdk_chemistry.utils import compute_valence_space_parameters
from qdk_chemistry.constants import ANGSTROM_TO_BOHR
from qdk.widgets import Circuit

# Prepare H2 structure and perform SCF calculation
coords_ang = np.array([[0.0, 0.0, 0.0], [0.0, 0.0, 0.7408]])
structure = Structure(coords_ang * ANGSTROM_TO_BOHR, [Element.H, Element.H])
scf_solver = create("scf_solver")
_, wfn_hf = scf_solver.run(structure, charge=0, spin_multiplicity=1, basis_or_guess="sto-3g")

# Select valence active space
num_val_e, num_val_o = compute_valence_space_parameters(wfn_hf, charge=0)
selector = create("active_space_selector", "qdk_valence", num_active_electrons=num_val_e, num_active_orbitals=num_val_o)
active_wfn = selector.run(wfn_hf)

# Construct Hamiltonian and run CASCI
hamiltonian_constructor = create("hamiltonian_constructor")
active_hamiltonian = hamiltonian_constructor.run(active_wfn.get_orbitals())
alpha_electrons, beta_electrons = active_wfn.get_active_num_electrons()
mc = create("multi_configuration_calculator")
_, wfn_cas = mc.run(active_hamiltonian, alpha_electrons, beta_electrons)

# Truncate to top 4 determinants
top_dets = wfn_cas.get_top_determinants(4)
pmc_calculator = create("projected_multi_configuration_calculator", "macis_pmc")
active_hamiltonian = hamiltonian_constructor.run(wfn_cas.get_orbitals())
_, wfn_state = pmc_calculator.run(active_hamiltonian, list(top_dets.keys()))

# Sparse isometry without binary encoding
state_prep = create("state_prep", "sparse_isometry")
sparse_isometry_circuit = state_prep.run(wfn_state)

# Sparse isometry with binary encoding
state_prep_be = create("state_prep", "sparse_isometry", binary_encoding=True)
binary_encoding_circuit = state_prep_be.run(wfn_state)

# Save circuit diagram and print logical resource estimates
result = binary_encoding_circuit.estimate()
print(result.logical_counts)
qsharp_circuit = binary_encoding_circuit.get_qsharp_circuit()
Circuit(qsharp_circuit)
\end{lstlisting}
\caption{End-to-end code samples using \ac{QDK} Chemistry: loading a molecular structure, computing the multi-configurational wavefunction, and generating state preparation circuits using the sparse isometry method with and without binary encoding.}
\label{fig:code-scf}
\end{figure}

\begin{figure}[H]
\centering
\includegraphics[width=0.7\textwidth]{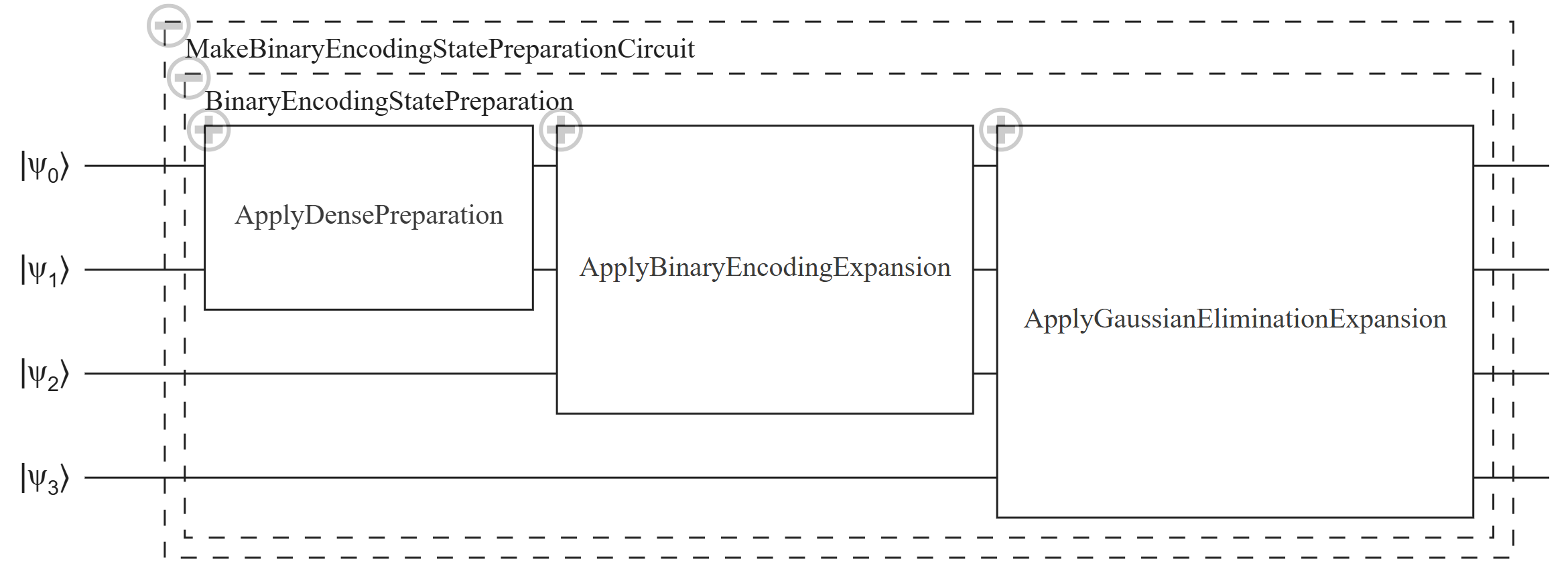}
\caption{Q\# circuit diagram for the sparse isometry state preparation with binary encoding. Each box represents a subroutine described in Section~\ref{sec:methods} and can be expanded in the Q\# visualizer to reveal the underlying quantum gates.}
\label{fig:sample-output-circuit}
\end{figure}

Running the script above produces the circuit diagram in Figure~\ref{fig:sample-output-circuit} and the following logical resource estimates:
{\footnotesize
\begin{verbatim}
{'numQubits': 4, 'tCount': 0, 'rotationCount': 3, 'rotationDepth': 2, 'cczCount': 0, 
'ccixCount': 0, 'measurementCount': 0}
\end{verbatim}
}

\end{document}